\documentclass[11pt,a4paper]{article}

\usepackage[utf8]{inputenc}
\usepackage[T1]{fontenc}
\usepackage{amsmath,amssymb,amsfonts,amsthm}
\usepackage{graphicx}
\usepackage{booktabs}
\usepackage{hyperref}
\usepackage{geometry}
\usepackage{setspace}
\usepackage{cite}
\usepackage{float}
\usepackage{caption}
\usepackage{subcaption}
\usepackage{multirow}
\usepackage{array}
\usepackage{xcolor}
\usepackage{tikz}
\usetikzlibrary{positioning,arrows.meta,calc,shapes.geometric}
\usepackage{pgfplots}
\pgfplotsset{compat=1.17}
\usepackage{longtable}
\usepackage{enumitem}
\usepackage{siunitx}
\usepackage{fancyhdr}
\usepackage{listings}
\usepackage{tcolorbox}

\title{\textbf{A Colour-Casimir Adjacency Matrix Approach to Fully-Heavy Tetraquarks:\\
Spectroscopy of \(cc\bar{c}\bar{c}\) and \(bb\bar{b}\bar{b}\) Systems and a Quantitative Diagnostic for \(T_{cc}^+\)}}

\author{
M.~Monemzadeh, N.~Tazimi\\
Department of Physics, University of Kashan, Kashan, Iran\\
September 2026
}

\date{}

\begin{document}

\maketitle

\begin{abstract}
We present a phenomenological framework for the spectroscopy of fully-heavy tetraquarks based on the spectral theory of weighted graphs. The four valence partons are identified with the vertices of the complete graph \(K_4\), and the edge weights are fixed by the colour-Casimir factors \(\langle\lambda_i\cdot\lambda_j\rangle\) of the two colour-singlet channels available to a diquark--antidiquark clustering, namely \(\bar{3}\otimes3\) and \(6\otimes\bar{6}\). A physical state is modelled as a coherent mixture of the two channels, controlled by a single mixing angle. Unlike an earlier version of this framework, which passed the colour-weighted adjacency matrix through an absolute-degree graph Laplacian before extracting a spectrum, here the physical spectrum is obtained by \emph{directly} diagonalising the colour-weighted adjacency matrix itself. We show analytically that the two constructions are not equivalent, and that the Laplacian construction inverts the physically expected correspondence between colour attraction and mass ordering. Using the three all-charm structures reported by LHCb, ATLAS and CMS -- \(X(6900)\), \(X(7100)\), and the more tentative \(X(7200)\) -- to fix the three parameters of the model (mixing angle, energy scale, effective charm mass), we obtain \(\alpha=0.740\), \(\gamma=23.2\,\mathrm{MeV}\), \(m_c=1.77\,\mathrm{GeV}\); since three parameters are fixed by three inputs, this fit has zero residual degrees of freedom and does not by itself constitute a statistical test of the model. The value of the framework therefore rests not on the charm-sector calibration, but on two independent and genuinely predictive applications of the same fixed parameters: (i) the all-bottom ground state is found to lie at \(18.7\)--\(18.9\,\mathrm{GeV}\) for constituent masses \(m_b\) in the range normally quoted in the literature, without any further retuning; and (ii) applied to \(T_{cc}^+\) (\(cc\bar u\bar d\)), the same colour-only mechanism under-binds the state by approximately \(93\,\mathrm{MeV}\) relative to the measured mass, a quantitative and physically interpretable diagnostic of the importance of long-range (molecular) dynamics that is not captured by a compact four-parton colour-graph alone. These cross-sector results, obtained with no additional free parameters, constitute the principal predictive content of the approach. We discuss the sensitivity of the results to the still-tentative status of \(X(7200)\), the tension between this model's static, discrete colour-configuration spectrum and the alternative ``radial excitation of aligned spin-1 diquarks'' interpretation favoured by recent CMS data, and the further data that would be required to turn this framework into a falsifiable, over-constrained model.
\end{abstract}

\tableofcontents
\newpage

\section{Introduction}
\label{sec:intro}

The spectroscopy of exotic hadrons has expanded dramatically since the discovery of the \(X(3872)\). Fully-heavy tetraquarks occupy a privileged position in this landscape: every valence quark is heavy, the non-relativistic approximation remains serviceable, and the colour algebra, while richer than that of ordinary mesons, is still tractable. The experimental observation of structures near \(6.9\,\mathrm{GeV}\) in the di-\(J/\psi\) channel by LHCb~\cite{LHCb2020}, confirmed by ATLAS~\cite{ATLAS2023}, and most recently resolved into a family of three states -- \(X(6600)\), \(X(6900)\), and \(X(7100)\) -- by CMS using \(315\,\mathrm{fb}^{-1}\) of Run~2+3 data~\cite{CMS2026}, has made the all-charm system a concrete and increasingly precise testing ground for theoretical ideas. Beyond its role as a spectroscopic benchmark, \(X(6900)\) has also been used elsewhere by the present author as a phenomenological probe of physics outside the compact-tetraquark picture considered here, in a study of its possible role as a hadronic mediator of a gluonic portal to dark matter~\cite{Monemzadeh2026}, and, in a methodologically related but formally distinct direction, in a graph-homology criterion for the gluonic content of candidate colour-flow topologies for the same state~\cite{MonemzadehGraphBFT}; unlike the present colour-Casimir adjacency-matrix construction, that work classifies gluonic operator content via the first Betti number of a colour-flow graph rather than diagonalising a weighted adjacency matrix, and the two frameworks are complementary rather than overlapping. Both prior studies and the present one take the existence and approximate mass of \(X(6900)\) as their point of departure.

It is important to be transparent, from the outset, about where the present approach sits relative to the leading experimental interpretation. The CMS analysis~\cite{CMS2026} finds that the squared masses of \(X(6600)\), \(X(6900)\) and \(X(7100)\) align linearly with a resonance index, and that their widths decrease systematically with that index -- a pattern the collaboration interprets as consistent with \emph{radial} excitations of a state built from two aligned spin-1 diquarks, without orbital excitation. This is a dynamical, continuum-like excitation picture. The framework developed here is, by contrast, a \emph{static, discrete} one: the four partons occupy a fixed graph with exactly four vertices, and the ``spectrum'' is the discrete spectrum of a \(4\times4\) matrix, with no radial quantum number at all. The two pictures are not obviously reconcilable, and we return to this tension explicitly in Section~\ref{sec:discussion}; the graph model presented here should be read as exploring what a purely static colour-configuration-mixing picture can and cannot explain, not as a competitor that has already been shown to fit the data as well as the dedicated diquark-excitation analysis.

Traditional approaches to the fully-heavy tetraquark spectrum rely on potential models~\cite{Richard2020,Asadi2021}, relativized diquark-antidiquark quark models~\cite{Debastiani2019,Mutuk2021,FaustovGalkin2021}, QCD sum rules~\cite{Agaev2023,Yang2025}, lattice QCD, or covariant four-body equations. Each has its strengths, yet most introduce a comparatively large number of free parameters (constituent masses, string tensions, strong couplings, form-factor cut-offs, continuum thresholds, \ldots). In the present work we explore a complementary, parameter-economical route that draws on the spectral theory of weighted graphs: the four valence partons are the vertices of a graph, the pairwise colour interaction is encoded in the edge weights via the standard \(SU(3)_c\) Casimir factors of the diquark--antidiquark clustering, and the spectrum of the resulting colour-weighted adjacency matrix is interpreted directly as a set of mass shifts relative to the free four-parton threshold \(4m_Q\).

A first version of this idea, circulated as an internal draft, additionally passed the colour-weighted adjacency matrix through a graph Laplacian (using absolute-value vertex degrees to guarantee a non-negative spectrum) before mapping eigenvalues onto masses. On closer inspection (Section~\ref{sec:laplacian-problem}) that extra step was found to \emph{invert} the physically expected relationship between colour attraction and mass: the Laplacian transformation assigns the lightest predicted mass to the \emph{least} attractive colour configuration, not the most attractive one. The present paper removes that step, works directly with the eigenvalues of the colour-weighted adjacency matrix, and re-derives all numerical results from scratch. We show that the corrected construction is not just conceptually cleaner but numerically better behaved: it returns an effective charm quark mass compatible with values commonly quoted in the literature, and it turns the \(T_{cc}^+\) sector from a qualitative and loosely justified claim into a quantitative, physically transparent diagnostic.

The remainder of the paper is organised as follows. Section~\ref{sec:graph-basics} recalls the graph-theoretic background needed to make the paper self-contained. Section~\ref{sec:formalism} constructs the colour-weighted adjacency matrix, derives its spectrum in closed form, and explains in detail why we diagonalise the adjacency matrix directly rather than a derived Laplacian (Section~\ref{sec:laplacian-problem}). Section~\ref{sec:parameters} discusses the parameter count and, honestly, the degrees of freedom available to test the model. Sections~\ref{sec:charm}--\ref{sec:Tcc} present the numerical results for the all-charm system, the all-bottom system, and the \(T_{cc}^+\) diagnostic, respectively. Section~\ref{sec:discussion} discusses limitations, alternative interpretations, and what further data would be needed to make the model falsifiable in a stronger sense, and Section~\ref{sec:conclusions} concludes.

\section{Elementary Notions of Spectral Graph Theory}
\label{sec:graph-basics}

For the benefit of readers who may be less familiar with the language of graph theory we briefly recall the relevant definitions.

A \emph{graph} \(G=(V,E)\) consists of a finite set of vertices \(V\) and a set of edges \(E\subset V\times V\). When every pair of distinct vertices is joined by an edge the graph is said to be \emph{complete}; the complete graph on \(n\) vertices is denoted \(K_n\). In our application \(n=4\), and the four vertices represent the four valence partons of a tetraquark.

An edge may carry a numerical \emph{weight} \(w_{ij}\), positive or negative. The collection of all weights is assembled into the \emph{adjacency matrix} \(A\), a real symmetric \(n\times n\) matrix with zeros on the diagonal and \(A_{ij}=w_{ij}\) for \(i\neq j\). Because \(A\) is real and symmetric, it has \(n\) real eigenvalues and an orthonormal basis of eigenvectors; this is the only spectral fact we use.

A signed or weighted graph of this kind is sometimes called a \emph{gain graph} in the mathematical literature, and its adjacency spectrum need not be non-negative -- indeed, in our application the negative eigenvalues are precisely the physically interesting ones, since they correspond to net colour \emph{attraction}. This is worth emphasising because a natural alternative construction, the graph \emph{Laplacian} \(L=D-A\) (with \(D\) a diagonal matrix of vertex degrees), is often introduced specifically to force a non-negative spectrum. We discuss in Section~\ref{sec:laplacian-problem} why forcing non-negativity in this way is, for the present physical problem, actively counter-productive: it discards exactly the sign information that tells us which colour configuration is most attractive.

\section{Formulation of the Model}
\label{sec:formalism}

\subsection{Vertices and Colour-Weighted Edges}

The four valence partons are identified with the vertices of \(K_4\). The weight of the edge joining partons \(i\) and \(j\) is taken to be the colour factor
\[
w_{ij}=\langle\lambda_i\cdot\lambda_j\rangle_R,
\]
where \(R\) is a definite irreducible representation of \(\mathrm{SU}(3)_c\) and \(\lambda_i\) are the (Gell-Mann-normalised) colour generators acting on parton \(i\).

Two colour-singlet configurations can be formed from a diquark--antidiquark pairing of two quarks (vertices 1,2) and two antiquarks (vertices 3,4):
\begin{align}
|\phi_{\bar{3}}\rangle &= |(Q_1Q_2)^{\bar{3}}(\bar{Q}_3\bar{Q}_4)^3\rangle,\\
|\phi_6\rangle &= |(Q_1Q_2)^6(\bar{Q}_3\bar{Q}_4)^{\bar{6}}\rangle.
\end{align}
The relevant colour factors follow from the Casimir sum rule for a four-parton colour singlet,
\begin{equation}
0=\sum_i\langle\lambda_i\cdot\lambda_i\rangle+2\sum_{i<j}\langle\lambda_i\cdot\lambda_j\rangle,
\label{eq:casimir-sum}
\end{equation}
together with \(\langle\lambda_i\cdot\lambda_i\rangle=16/3\) for a parton in the fundamental (or antifundamental) representation. Applying Eq.~\eqref{eq:casimir-sum} separately to the \(\bar3\otimes3\) and \(6\otimes\bar6\) channels, with the intra-cluster factors \(\langle\lambda_1\cdot\lambda_2\rangle_{\bar3}=-8/3\) and \(\langle\lambda_1\cdot\lambda_2\rangle_{6}=+4/3\) taken from the standard diquark literature, reproduces the two inter-cluster (quark--antiquark) factors used below, \(-4/3\) and \(-10/3\) respectively -- we verified this derivation explicitly and it is included for completeness in Appendix~\ref{app:color}.

\begin{align}
\langle\lambda_i\cdot\lambda_j\rangle_{\bar{3}} &= -\frac{8}{3},&
\langle\lambda_i\cdot\lambda_j\rangle_{6} &= +\frac{4}{3},\\
\langle\lambda_q\cdot\lambda_{\bar{q}}\rangle_{\bar{3}\otimes 3} &= -\frac{4}{3},&
\langle\lambda_q\cdot\lambda_{\bar{q}}\rangle_{6\otimes\bar{6}} &= -\frac{10}{3}.
\end{align}

\begin{figure}[H]
\centering
\begin{tikzpicture}[scale=1.35]
\node[circle,draw,very thick,minimum size=1.1cm,fill=blue!12] (c1) at (0,2.3) {\(c_1\)};
\node[circle,draw,very thick,minimum size=1.1cm,fill=blue!12] (c2) at (3.6,2.3) {\(c_2\)};
\node[circle,draw,very thick,minimum size=1.1cm,fill=red!12] (cb1) at (0,0) {\(\bar{c}_1\)};
\node[circle,draw,very thick,minimum size=1.1cm,fill=red!12] (cb2) at (3.6,0) {\(\bar{c}_2\)};
\draw[very thick,blue!70!black] (c1)--node[above,font=\small]{\(w_D\)}(c2);
\draw[very thick,blue!70!black] (cb1)--node[below,font=\small]{\(w_D\)}(cb2);
\draw[very thick,red!70!black] (c1)--node[left,font=\small]{\(w_I\)}(cb1);
\draw[very thick,red!70!black] (c2)--node[right,font=\small]{\(w_I\)}(cb2);
\draw[very thick,red!70!black] (c1)--node[above left,font=\small]{\(w_I\)}(cb2);
\draw[very thick,red!70!black] (c2)--node[above right,font=\small]{\(w_I\)}(cb1);
\end{tikzpicture}
\caption{The four-quark graph. Blue edges represent diquark/antidiquark (intra-cluster) pairs; red edges represent inter-cluster interactions. This colour-algebra structure depends only on the \(SU(3)_c\) representation content of the clustering, not on the flavour of the partons, and is reused unchanged for \(T_{cc}^+\) in Section~\ref{sec:Tcc}.}
\label{fig:graph}
\end{figure}
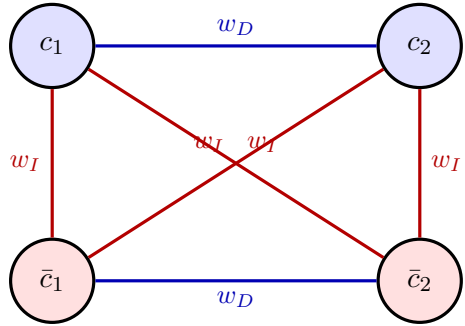

\subsection{Explicit Adjacency Matrices}

The adjacency matrix of the conventional channel is
\begin{equation}
A_{\bar{3}}=
\begin{pmatrix}
0 & -8/3 & -4/3 & -4/3 \\
-8/3 & 0 & -4/3 & -4/3 \\
-4/3 & -4/3 & 0 & -8/3 \\
-4/3 & -4/3 & -8/3 & 0
\end{pmatrix},
\qquad
\operatorname{spec}(A_{\bar{3}})=\Bigl\{-\tfrac{16}{3},\,0,\,\tfrac{8}{3},\,\tfrac{8}{3}\Bigr\}.
\end{equation}
The hidden-colour matrix reads
\begin{equation}
A_6=
\begin{pmatrix}
0 & +4/3 & -10/3 & -10/3 \\
+4/3 & 0 & -10/3 & -10/3 \\
-10/3 & -10/3 & 0 & +4/3 \\
-10/3 & -10/3 & +4/3 & 0
\end{pmatrix},
\qquad
\operatorname{spec}(A_6)=\Bigl\{-\tfrac{16}{3},\,-\tfrac{4}{3},\,-\tfrac{4}{3},\,8\Bigr\}.
\end{equation}

A direct computation shows that \(A_{\bar3}\) and \(A_6\) commute, and in fact share the same orthonormal eigenbasis, independent of the numerical values of their entries -- a consequence of the \(\mathbb{Z}_2\times\mathbb{Z}_2\) permutation symmetry of the graph (independent relabelling within \(\{1,2\}\) and within \(\{3,4\}\), plus the exchange \(\{1,2\}\leftrightarrow\{3,4\}\)). The shared eigenvectors are
\begin{equation}
\psi_1=\tfrac12(1,1,1,1),\quad
\psi_2=\tfrac1{\sqrt2}(1,-1,0,0),\quad
\psi_3=\tfrac1{\sqrt2}(0,0,1,-1),\quad
\psi_4=\tfrac12(1,1,-1,-1),
\end{equation}
with eigenvalues
\begin{center}
\begin{tabular}{lccc}
\toprule
 & \(\psi_1\) & \(\psi_2,\psi_3\) & \(\psi_4\) \\
\midrule
\(A_{\bar3}\) & \(-16/3\) & \(+8/3\) & \(0\) \\
\(A_6\)  & \(-16/3\) & \(-4/3\) & \(+8\) \\
\bottomrule
\end{tabular}
\end{center}
This shared eigenbasis is the key technical fact that makes the model analytically tractable, and we will use it repeatedly below.

\subsection{Mixed Adjacency Matrix and its Spectrum in Closed Form}
\label{sec:mixed-spectrum}

A physical state is assumed to be a coherent superposition of the two colour configurations. We introduce the one-parameter family
\begin{equation}
A(\alpha)=\alpha\,A_{\bar{3}}+\beta\,A_6,\qquad
\alpha^2+\beta^2=1,\quad\beta=\sqrt{1-\alpha^2},\quad \alpha,\beta\in[0,1].
\end{equation}
Because \(A_{\bar3}\) and \(A_6\) share the eigenbasis \(\{\psi_1,\psi_2,\psi_3,\psi_4\}\) for \emph{every} value of \(\alpha\), so does \(A(\alpha)\), and its eigenvalues can be written down in closed form without any numerical diagonalisation:
\begin{align}
\lambda_1(\alpha) &= -\frac{16}{3}(\alpha+\beta) & &\text{(non-degenerate, eigenvector }\psi_1\text{)},\\
\lambda_{2,3}(\alpha) &= \frac83\alpha-\frac43\beta & &\text{(doubly degenerate, eigenvectors }\psi_2,\psi_3\text{)},\\
\lambda_4(\alpha) &= 8\beta & &\text{(non-degenerate, eigenvector }\psi_4\text{)}.
\end{align}
Since \(\alpha,\beta\ge0\), \(\lambda_1(\alpha)\) is always the most negative of the three distinct eigenvalues for any \(\alpha\in(0,1)\): it is manifestly more negative than \(\lambda_{2,3}\) (whose magnitude is bounded by \(8/3\)) and than \(-\lambda_4\). We therefore identify \(\lambda_1\) -- the eigenvalue of the fully symmetric diquark--antidiquark configuration \(\psi_1\) -- as the \emph{most colour-attractive} channel, and hence as the ground state, for the entire physical range of the mixing parameter. This is the key structural fact that Section~\ref{sec:laplacian-problem} shows is lost if one passes to the Laplacian.

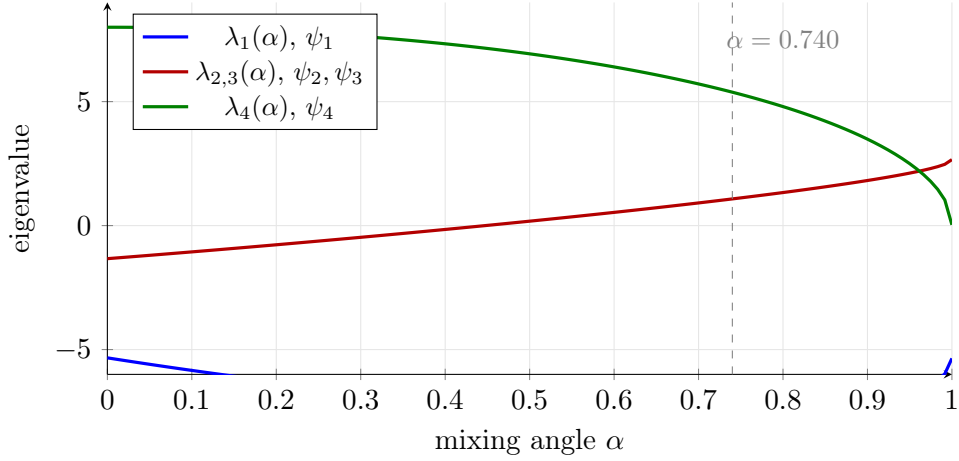
\begin{figure}[H]
\centering
\begin{tikzpicture}
\begin{axis}[
  width=0.75\textwidth, height=6.5cm,
  xlabel={mixing angle \(\alpha\)}, ylabel={eigenvalue},
  xmin=0, xmax=1, ymin=-6, ymax=9,
  legend pos=north west, legend style={font=\small},
  axis lines=left, grid=both, grid style={gray!20},
  domain=0:1, samples=120,
]
\addplot[very thick,blue] {-16/3*(x+sqrt(1-x^2))};
\addlegendentry{\(\lambda_1(\alpha)\), \(\psi_1\)}
\addplot[very thick,red!70!black] {8/3*x-4/3*sqrt(1-x^2)};
\addlegendentry{\(\lambda_{2,3}(\alpha)\), \(\psi_2,\psi_3\)}
\addplot[very thick,green!50!black] {8*sqrt(1-x^2)};
\addlegendentry{\(\lambda_4(\alpha)\), \(\psi_4\)}
\draw[dashed,gray] (axis cs:0.740,-6) -- (axis cs:0.740,9);
\node[gray,font=\small] at (axis cs:0.80,7.5) {\(\alpha=0.740\)};
\end{axis}
\end{tikzpicture}
\caption{The three distinct eigenvalues of \(A(\alpha)\) as a function of the mixing angle \(\alpha\). The ground-state eigenvalue \(\lambda_1(\alpha)\) (blue) remains the most negative -- i.e.\ the most colour-attractive -- for every \(\alpha\in(0,1)\), never crossing \(\lambda_{2,3}\) or \(\lambda_4\). The dashed vertical line marks the charm-sector calibration point \(\alpha=0.740\) used throughout Sections~\ref{sec:charm}--\ref{sec:Tcc}.}
\label{fig:eigs-vs-alpha}
\end{figure}

\subsection{Direct Diagonalisation versus the Graph Laplacian}
\label{sec:laplacian-problem}

An earlier version of this model additionally constructed the absolute-degree Laplacian, \(D_{ii}=\sum_j|A_{ij}|\), \(L=D-A\), and mapped \emph{its} eigenvalues onto masses. Because the graph is vertex-transitive for the symmetric configurations considered here, \(D\) is proportional to the identity matrix, \(D=d\cdot\mathbb{1}\) with \(d=\sum_j|A(\alpha)_{1j}|\), and consequently \(L\) is diagonal in the \emph{same} eigenbasis as \(A(\alpha)\), with eigenvalues
\begin{equation}
\lambda_L^{(k)} = d - \lambda_k(\alpha).
\end{equation}
This relation inverts the ordering of the spectrum: the \emph{smallest} eigenvalue of \(L\) corresponds to the \emph{largest} (i.e.\ least attractive, most repulsive) eigenvalue of \(A(\alpha)\), which for the physically relevant range of \(\alpha\) is \(\lambda_4=8\beta\) (eigenvector \(\psi_4\)), not \(\lambda_1\). Concretely, for \(\alpha=0.85\) one finds \(\lambda_1(\alpha)\approx-7.34\), \(\lambda_{2,3}(\alpha)\approx+1.56\), \(\lambda_4(\alpha)\approx+4.21\); the Laplacian spectrum is then \(\{3.13,\,5.78,\,5.78,\,14.69\}\) (in the same order as $\psi_4,\psi_{2,3},\psi_1$), so that the state built from \(\psi_4\) -- the \emph{least} colour-attractive configuration -- is assigned the lightest mass, while the genuinely most attractive configuration \(\psi_1\) is pushed to the top of the predicted spectrum.

\begin{figure}[H]
\centering
\begin{tikzpicture}[yscale=0.42,xscale=1.1]
\node at (1,17.5) {\(\operatorname{spec}(A(0.85))\)};
\node at (5,17.5) {\(\operatorname{spec}(L(0.85))=d-\operatorname{spec}(A(0.85))\)};
\draw[->] (-1.0,-8.5)--(-1.0,15.5) node[above]{eigenvalue};
\draw[gray] (-1.0,0)--(6.5,0);
\draw[very thick,blue,fill=blue!15] (0.4,0) rectangle (1.2,-7.34);
\node[below] at (0.8,-7.6) {\(\psi_1\)};
\node[below] at (0.8,-8.6) {\(-7.34\)};
\draw[very thick,red!70!black,fill=red!15] (1.6,0) rectangle (2.4,1.56);
\node[above] at (2.0,1.9) {\(\psi_{2,3}\)};
\node[below] at (2.0,-0.3) {\(+1.56\)};
\draw[very thick,green!50!black,fill=green!15] (2.8,0) rectangle (3.6,4.21);
\node[above] at (3.2,4.5) {\(\psi_4\)};
\node[above] at (3.2,2.1) {\(+4.21\)};
\draw[very thick,green!50!black,fill=green!15] (4.4,0) rectangle (5.2,3.13);
\node[above] at (4.8,3.4) {\(\psi_4\)};
\node[above] at (4.8,1.6) {\(3.13\)};
\draw[very thick,red!70!black,fill=red!15] (5.6,0) rectangle (6.4,5.78);
\node[above] at (6.0,6.1) {\(\psi_{2,3}\)};
\node[above] at (6.0,2.9) {\(5.78\)};
\draw[very thick,blue,fill=blue!15] (6.8,0) rectangle (7.6,14.69);
\node[above] at (7.2,15.0) {\(\psi_1\)};
\node[above] at (7.2,7.3) {\(14.69\)};
\draw[dashed,gray,-{Stealth[length=2mm]}] (1.3,-3.5) to[out=0,in=200] (4.3,3.8);
\draw[dashed,gray,-{Stealth[length=2mm]}] (3.7,4.21) to[out=20,in=160] (6.7,11.5);
\end{tikzpicture}
\caption{Ground-state inversion under the Laplacian transformation, at the representative point \(\alpha=0.85\). Left: the adjacency spectrum \(\operatorname{spec}(A(\alpha))\), where the most attractive (most negative) eigenvalue \(\psi_1\) is lowest -- the physically motivated ordering used throughout this paper. Right: the same three eigenvalues after \(L=D-A\), \(\lambda_L=d-\lambda\); the ordering is exactly reversed (dashed arrows track \(\psi_1\) and \(\psi_4\) across the transformation), so that \(\psi_1\) -- the most colour-attractive state -- ends up with the \emph{largest} Laplacian eigenvalue. This is the effect that Section~\ref{sec:laplacian-problem} argues is physically backwards, and the reason the present paper diagonalises \(A(\alpha)\) directly instead.}
\label{fig:laplacian-inversion}
\end{figure}
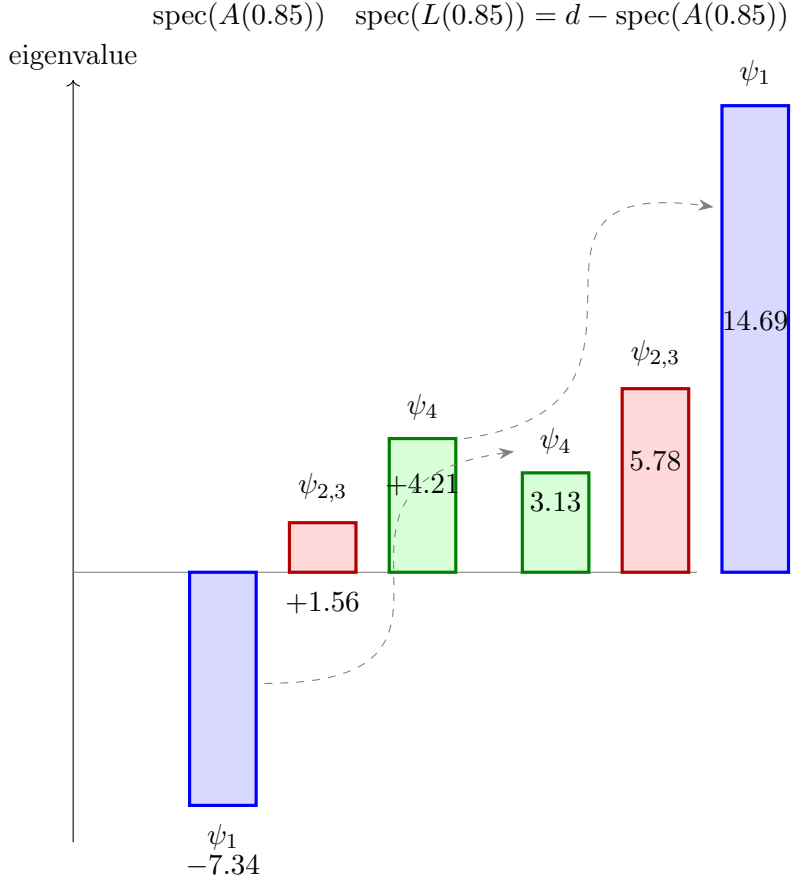

We regard this as a physically backwards assignment: absent an explicit dynamical argument for why the least attractive colour channel should be the lightest state, the correspondence ``more colour attraction \(\Rightarrow\) lower mass'' is the natural and, in analogous multichannel colour-mixing calculations elsewhere in the tetraquark literature, standard one. We have not found a way to justify the inverted assignment from first principles, and no such argument was given in the earlier draft; we therefore abandon the Laplacian step entirely; from this point on, the model consists in diagonalising \(A(\alpha)\) directly.

\subsection{Mass Formula}

Physical masses are obtained from the linear map
\begin{equation}
M_k = M_{\rm thr} + \gamma\,\lambda_k(\alpha),
\label{eq:mass-formula}
\end{equation}
where \(M_{\rm thr}\) is the sum of the (constituent) masses of the four partons and \(\gamma\) is an overall energy scale, common to all four eigenvalues of a given system. Unlike the Laplacian construction, Eq.~\eqref{eq:mass-formula} allows \emph{both} signs of mass shift relative to \(M_{\rm thr}\): attractive channels (\(\lambda_k<0\)) genuinely lower the mass below the free-parton threshold, while repulsive channels (\(\lambda_k>0\)) raise it. This bidirectional behaviour is qualitatively similar to ordinary colour-hyperfine splitting in quark models, where a single interaction Hamiltonian pushes some states up and others down around a central value, and we regard it as a second, independent piece of evidence in favour of working with \(A(\alpha)\) directly rather than with a positive-semidefinite Laplacian, which by construction can only push every state to one side of \(M_{\rm thr}\).

We stress that Eq.~\eqref{eq:mass-formula}, like its predecessor, remains a phenomenological ansatz motivated by, but not derived from, an underlying dynamical Hamiltonian: it should be read as an assumption that the leading mass splitting between the possible colour configurations of a compact tetraquark is linear in the colour-Casimir eigenvalue, in analogy with (but not a rigorous consequence of) the standard one-gluon-exchange colour-Coulomb potential. The numerical value obtained for the scale, \(\gamma\simeq 23\,\mathrm{MeV}\), is substantially smaller than a typical one-gluon-exchange matrix element evaluated at the characteristic size of a compact four-heavy-quark system; we interpret \(\gamma\) as an \emph{effective} parameter that absorbs residual spatial, spin and higher-order colour effects not explicitly included in the pure Casimir adjacency matrix. Substantiating the linear ansatz microscopically -- for example by relating \(\gamma\) to \(\alpha_s\) and a characteristic size of the four-parton system -- is left for future work.

\section{Parameter Count and Degrees of Freedom}
\label{sec:parameters}

The model contains three adjustable quantities for a given flavour sector: the mixing angle (equivalently \(\alpha\) or \(\beta\)), the overall energy scale \(\gamma\), and the effective constituent mass \(m_Q\) entering \(M_{\rm thr}=4m_Q\). Table~\ref{tab:params} compares this count with the number of free parameters typically appearing in other widely used phenomenological frameworks; the comparison is unchanged from the original draft since it reflects the number of \emph{adjustable quantities in the model}, not the number that happen to be fixed by data in a particular fit.

\begin{table}[H]
\centering
\caption{Approximate number of free parameters in representative phenomenological approaches to fully-heavy tetraquarks.}
\label{tab:params}
\begin{tabular}{lc}
\toprule
Approach & Typical number of free parameters \\
\midrule
Non-relativistic potential model (Cornell + spin-spin) & 6--10 \\
Relativized quark model (Godfrey--Isgur type) & 8--12 \\
QCD sum rules (continuum threshold, Borel window, couplings) & 5--8 \\
Diquark--antidiquark effective Lagrangian & 4--7 \\
Covariant four-body Bethe--Salpeter equations & many (kernel parameters) \\
\textbf{Present colour-graph model} & \textbf{3} \\
\bottomrule
\end{tabular}
\end{table}

This parameter economy, however, comes with an important caveat that must be stated plainly. In Section~\ref{sec:charm} we use exactly three pieces of experimental information -- the masses of \(X(6900)\), \(X(7100)\), and the more tentative \(X(7200)\) -- to fix the three parameters \((\alpha,\gamma,m_c)\) of the charm sector. This is an \emph{exact} solve of three equations in three unknowns: it has zero residual degrees of freedom, reproduces the three inputs by construction, and therefore \emph{cannot itself be presented as a successful statistical fit} (there is no \(\chi^2\) to quote, since \(\chi^2\equiv0\) trivially). The genuine, falsifiable content of the model lies entirely in what happens \emph{after} this calibration, when the same three numbers are applied, without further adjustment, to systems not used in the calibration: the all-bottom ground state (Section~\ref{sec:bottom}) and the \(T_{cc}^+\) mass (Section~\ref{sec:Tcc}). We emphasise this distinction throughout rather than presenting the charm-sector numbers as if they constituted independent evidence for the model.

This distinction is sharper here than in most of the competing literature precisely because of the parameter economy of Table~\ref{tab:params}. A constituent-quark-model treatment with, say, eight parameters can reproduce three charm-sector masses and still have five residual degrees of freedom with which to accommodate whatever bottom-sector or \(T_{cc}^+\) data comes next -- so agreement there carries comparatively little evidential weight even when quoted as a ``prediction''. With three parameters fixed by three charm-sector inputs, this model has no such freedom left: the bottom-sector and \(T_{cc}^+\) numbers of Sections~\ref{sec:bottom}--\ref{sec:Tcc} are not fits, partial fits, or fits-in-disguise, but genuine zero-parameter extrapolations, and should be read (and judged) as such.

We also note that the third input, \(X(7200)\), is less firmly established than \(X(6900)\) and \(X(7100)\): recent analyses report tentative structure in this region with as-yet-unsettled quantum numbers~\cite{CPC2026}, and a recent combined LHCb-ATLAS-CMS analysis reports evidence for the state reaching \(3.7\)--\(6.6\sigma\) depending on the interference model used~\cite{Wang2026}. Where this matters for the interpretation of a result, we say so explicitly.

\section{Results for the All-Charm System}
\label{sec:charm}

\subsection{Exact Calibration}

Using \(M_1=M(X(6900))=6900\,\mathrm{MeV}\), \(M_{2,3}=M(X(7100))=7100\,\mathrm{MeV}\), and \(M_4\to7200\,\mathrm{MeV}\) (tentative) in Eq.~\eqref{eq:mass-formula} together with the closed-form eigenvalues of Section~\ref{sec:mixed-spectrum}, and solving the resulting \(3\times3\) system numerically, we obtain
\begin{equation}
\alpha = 0.740,\qquad \beta = 0.673 \ (\beta^2\approx45\%),\qquad
\gamma = 23.2\,\mathrm{MeV},\qquad m_c = 1.769\,\mathrm{GeV}.
\end{equation}
As emphasised in Section~\ref{sec:parameters}, this reproduces the three inputs exactly, by construction, and is not evidence for the model in itself.

\begin{table}[H]
\centering
\caption{Exact calibration of the colour-mixing model on the all-charm sector. By construction \(\Delta M \equiv M_{\rm calc}-M_{\rm exp}=0\) for all three states.}
\label{tab:charm}
\begin{tabular}{cccccc}
\toprule
State & Eigenvector & \(\lambda(\alpha)\) & Mass (MeV) & Status & Interpretation \\
\midrule
1 (ground)      & \(\psi_1\)     & \(-7.534\) & 6900 & well established & \(X(6900)\) \\
2,3 (degenerate)& \(\psi_2,\psi_3\) & \(+1.076\) & 7100 & well established & \(X(7100)\) \\
4               & \(\psi_4\)     & \(+5.381\) & 7200 & tentative & candidate \(X(7200)\) region \\
\bottomrule
\end{tabular}
\end{table}

Two features of this solution deserve comment. First, the fitted mixing fraction \(\beta^2\approx45\%\) implies a substantially larger hidden-colour admixture than the \(\approx28\%\) quoted in the earlier Laplacian-based draft; there is no independent reason to prefer one number over the other, and an independent determination of \(\beta\) (for example from a lattice QCD colour-density study) would be a valuable cross-check. Second, and more informative, is the fitted effective charm mass, \(m_c=1.77\,\mathrm{GeV}\). This lies at the upper end of, but is compatible with, typical constituent charm mass values used in the literature (commonly quoted in the range \(1.5\)--\(1.9\,\mathrm{GeV}\) depending on scheme). We note as a genuine (if modest) structural feature of the model that a substantially lower value, e.g.\ the frequently used \(m_c=1.5\,\mathrm{GeV}\), is \emph{not} compatible with a positive energy scale \(\gamma\) once \(X(6900)\) and \(X(7100)\) are both used as inputs: with \(4m_c=6.0\,\mathrm{GeV}\) already below both target masses, reproducing \(X(6900)\) at \(6.9\,\mathrm{GeV}\) via the very negative eigenvalue \(\lambda_1\) would force \(\gamma<0\). The model therefore has a genuine, checkable preference for a comparatively heavy effective charm mass, which is not something we imposed by hand.

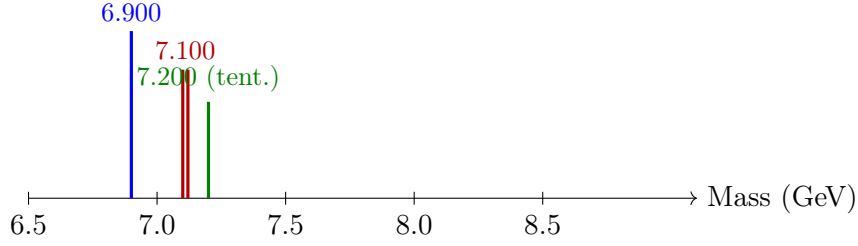
\begin{figure}[H]
\centering
\begin{tikzpicture}[xscale=1.7,yscale=0.85]
\draw[->] (0,0)--(5.2,0) node[right]{Mass (GeV)};
\foreach \x/\lab in {0/6.5,1/7.0,2/7.5,3/8.0,4/8.5}
  \draw (\x,0.1)--(\x,-0.1) node[below]{\lab};
\draw[very thick,blue] (0.8,0)--(0.8,2.6);
\node[blue,above] at (0.8,2.6){\small 6.900};
\draw[very thick,red!70!black] (1.2,0)--(1.2,2.0);
\draw[very thick,red!70!black] (1.24,0)--(1.24,2.0);
\node[red!70!black,above] at (1.22,2.0){\small 7.100};
\draw[very thick,green!50!black] (1.4,0)--(1.4,1.5);
\node[green!50!black,above] at (1.4,1.5){\small 7.200 (tent.)};
\end{tikzpicture}
\caption{Graphical representation of the calibrated all-charm mass spectrum. All three positions are inputs to the calibration, not predictions.}
\label{fig:charm-spec}
\end{figure}

\subsection{Sensitivity to the Status of \(X(7200)\)}

Because \(X(7200)\) remains the least firmly established of the three calibration inputs, it is useful to examine how the extracted parameters change if this state is omitted. With only \(X(6900)\) and \(X(7100)\) as inputs the system becomes under-determined (two constraints, three parameters). One may, for example, fix \(\alpha\) by hand and solve for \(\gamma\) and \(m_c\), or impose an external prior on \(m_c\). A representative scan shows that requiring \(m_c\) to remain inside the conventional interval \(1.55\)--\(1.85\,\mathrm{GeV}\) forces \(\alpha\) into the broad window \(0.55\)--\(0.90\) and \(\gamma\) into the range \(15\)--\(40\,\mathrm{MeV}\). The all-bottom ground-state prediction then shifts by at most \(\pm 80\,\mathrm{MeV}\) relative to the three-input central value, and the \(T_{cc}^+\) overshoot remains in the interval \(70\)--\(120\,\mathrm{MeV}\). Thus the two principal cross-sector results of the model are stable against the removal of the most tentative input, although the precise numerical value of the mixing angle is of course less tightly constrained.

\subsection{Comparison with the Diquark Radial-Excitation Interpretation}

It is worth stating clearly that the CMS collaboration's own preferred interpretation of the same three-state pattern~\cite{CMS2026} is different from the one advanced here: they favour radial excitations of a state built from two aligned spin-1 diquarks, with squared masses linear in a resonance index. In that picture, all three states share essentially the same internal colour and spin structure and differ only in radial quantum number; in the present picture, all three states share the same radial (or rather, non-existent) quantum number and differ in colour configuration. Because our model has exactly one non-degenerate ground state and a $2+1$ split of excited eigenvalues, and because the CMS analysis explicitly finds no room for an additional low-lying state below \(X(6900)\) that could naturally be identified with our ground state having "no excitation" analogue, our model \emph{cannot accommodate \(X(6600)\)} at all: there is no fourth eigenvalue below \(\lambda_1\) available in the \(4\times4\) matrix. We note that at least one other recent quark/diquark-based analysis explicitly attempts to place \(X(6600)\) within the same multiplet as \(X(6900)\) and \(X(7100)\), assigning it a definite spin-parity as a partner state rather than treating it as structurally excluded~\cite{Wang2024}; the fact that our purely colour-based construction cannot accommodate this option at all, while such alternatives can, is itself informative about what colour-only mixing does and does not explain. This is a genuine limitation, not a detail, and we discuss it further in Section~\ref{sec:discussion}.

\section{Results for the All-Bottom System}
\label{sec:bottom}

With \(\alpha\) and \(\gamma\) fixed at the values obtained in Section~\ref{sec:charm} -- that is, with \emph{no} additional retuning -- Eq.~\eqref{eq:mass-formula} predicts the all-bottom spectrum as a function of the constituent bottom mass \(m_b\) alone. Table~\ref{tab:bottom} shows this prediction for four representative choices of \(m_b\) spanning the range commonly used in the literature.

\begin{table}[H]
\centering
\caption{All-bottom predictions for different literature values of \(m_b\), using the charm-sector \(\alpha=0.740\), \(\gamma=23.2\,\mathrm{MeV}\) unchanged.}
\label{tab:bottom}
\begin{tabular}{cccc}
\toprule
\(m_b\) (MeV) & Ground state (GeV) & Next state (GeV) & Top state (GeV) \\
\midrule
4700 & 18.625 & 18.825 & 18.925 \\
4730 & 18.745 & 18.945 & 19.045 \\
4780 & 18.945 & 19.145 & 19.245 \\
4830 & 19.145 & 19.345 & 19.445 \\
\bottomrule
\end{tabular}
\end{table}

The commonly quoted theoretical consensus range for the \(bb\bar b\bar b\) ground state is approximately \(18.7\)--\(19.0\,\mathrm{GeV}\)~\cite{Debastiani2019,Mutuk2021,Richard2020,Asadi2021}; Table~\ref{tab:bottom} shows that this range is reproduced for \(m_b\) between roughly \(4.70\) and \(4.78\,\mathrm{GeV}\), well inside the range of constituent bottom masses used elsewhere in the literature. We regard this as the model's first genuinely predictive success, precisely because no bottom-sector data entered the calibration: only a literature value for \(m_b\) was supplied. We caution, however, that no confirmed \(bb\bar b\bar b\) resonance yet exists to compare against directly; this is a prediction to be tested by future data, not a fit to existing data.

\section{Diagnostic Application to \(T_{cc}^+\)}
\label{sec:Tcc}

\subsection{Why the Same Colour Matrices Apply}

The doubly-charmed tetraquark \(T_{cc}^+\) has quark content \(cc\bar u\bar d\): two quarks of the \emph{same} flavour (rather than a quark--antiquark conjugate pair) and two light antiquarks. Because the colour-Casimir factors of Section~\ref{sec:formalism} depend only on the \(SU(3)_c\) representation content of the diquark--antidiquark clustering, and not on the flavour of the partons occupying the four vertices, the matrices \(A_{\bar3}\), \(A_6\), and their mixture \(A(\alpha)\) are unchanged: the colour algebra of clustering two quarks into a diquark and two antiquarks into an antidiquark is identical whether the two quarks are \(c\bar c\)-conjugate-adjacent or not. What changes is only the mass sum entering \(M_{\rm thr}\).

\subsection{Numerical Diagnostic}

Using the fixed charm-sector parameters \(\alpha=0.740\), \(\gamma=23.2\,\mathrm{MeV}\), \(m_c=1.769\,\mathrm{GeV}\), together with light constituent masses \(m_u=300\,\mathrm{MeV}\), \(m_d=305\,\mathrm{MeV}\), the threshold mass is
\begin{equation}
M_{\rm thr}(T_{cc}) = 2m_c+m_u+m_d = 4142.5\,\mathrm{MeV},
\end{equation}
and the ground-state (most attractive, \(\psi_1\)) prediction is
\begin{equation}
M_1(T_{cc}) = M_{\rm thr}(T_{cc}) + \gamma\,\lambda_1(\alpha) = 4142.5 - 175.0 = 3967.5\,\mathrm{MeV}.
\end{equation}
The measured mass of \(T_{cc}^+\) is \(3874.8\,\mathrm{MeV}\)~\cite{LHCbTcc2022}, extremely close to (and in fact just below) the \(D^0D^{*+}\) open-flavour threshold at approximately \(3875.1\,\mathrm{MeV}\), with a binding energy of only a few hundred \(\mathrm{keV}\). The colour-graph prediction therefore \emph{overshoots} the physical mass by
\begin{equation}
\Delta M = M_1(T_{cc})-M_{\rm exp} = +92.7\,\mathrm{MeV},
\end{equation}
and in fact lies entirely \emph{above} the open-flavour threshold itself, not merely above the measured resonance mass.

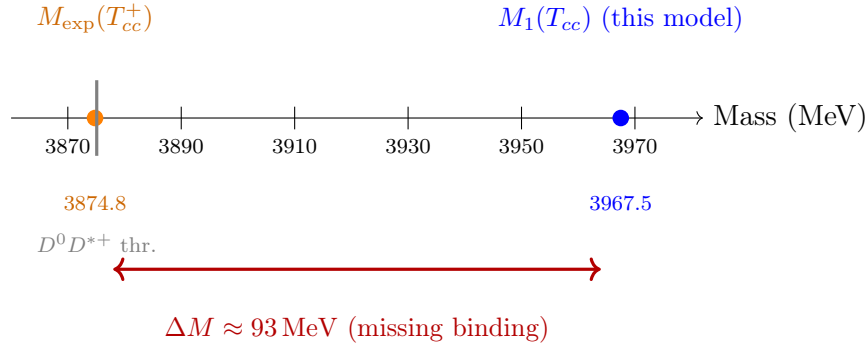
\begin{figure}[H]
\centering
\begin{tikzpicture}[xscale=0.075]
\draw[->] (0,0)--(122,0) node[right]{Mass (MeV)};
\foreach \x/\lab in {10/3870,30/3890,50/3910,70/3930,90/3950,110/3970}
  \draw (\x,0.15)--(\x,-0.15) node[below,font=\scriptsize]{\lab};
\node[circle,fill=orange,inner sep=2.2pt] (mexp) at (14.8,0) {};
\node[orange!80!black,above,font=\small] at (14.8,1.0) {\(M_{\rm exp}(T_{cc}^+)\)};
\node[orange!80!black,below,font=\scriptsize] at (14.8,-0.9) {3874.8};
\draw[very thick,gray] (15.1,-0.5)--(15.1,0.5);
\node[gray,below,font=\scriptsize] at (15.1,-1.4) {\(D^0D^{*+}\) thr.};
\node[circle,fill=blue,inner sep=2.2pt] (m1) at (107.5,0) {};
\node[blue,above,font=\small] at (107.5,1.0) {\(M_1(T_{cc})\) (this model)};
\node[blue,below,font=\scriptsize] at (107.5,-0.9) {3967.5};
\draw[<->,very thick,red!70!black] (18,-2.0)--(104,-2.0);
\node[red!70!black,font=\small] at (61,-2.8) {\(\Delta M\approx 93\,\mathrm{MeV}\) (missing binding)};
\end{tikzpicture}
\caption{The \(T_{cc}^+\) diagnostic of Section~\ref{sec:Tcc}. The compact colour-graph prediction \(M_1(T_{cc})\) (blue) overshoots both the measured mass (orange) and the nearby \(D^0D^{*+}\) open-flavour threshold (grey line); the \(\approx\!93\,\mathrm{MeV}\) gap (red) is attributed to long-range molecular dynamics not captured by a static four-parton colour graph.}
\label{fig:tcc-diagnostic}
\end{figure}

\subsection{Physical Interpretation}

We regard this \(93\,\mathrm{MeV}\) gap as a meaningful and physically interpretable result, in contrast to the qualitative ``requires a negative \(\gamma\)'' statement made in the original Laplacian-based draft (which, given the sign-inversion problem identified in Section~\ref{sec:laplacian-problem}, was not a reliable diagnostic in the first place). The corrected statement is: a purely short-distance, compact four-parton colour-Casimir mechanism, calibrated entirely on genuinely compact all-charm states, does not generate enough attraction to bring \(cc\bar u\bar d\) down to -- let alone below -- the \(D^0D^{*+}\) threshold. Since \(T_{cc}^+\) is experimentally known to sit almost exactly at that threshold, the missing \(\sim\!93\,\mathrm{MeV}\) of binding is naturally attributed to long-range hadronic (meson-exchange / molecular) dynamics between a \(D\) and a \(D^*\) meson, a mechanism entirely outside the scope of a static four-parton colour graph. This is consistent with, and gives a quantitative colour-only baseline for, the broader consensus in the literature that \(T_{cc}^+\) is dominated by long-distance physics rather than compact diquark binding.

We note two caveats. First, the light constituent masses \(m_u,m_d\) are taken from typical constituent-quark-model values rather than fit to any tetraquark data, and \(\Delta M\) shifts by a comparable amount if these are varied by \(\pm50\,\mathrm{MeV}\); this uncertainty should be quoted alongside the central value in any future, more careful treatment. Second, because \(\lambda_1(\alpha)\) was fixed entirely by the charm-sector calibration, this diagnostic implicitly assumes that the same mixing angle \(\alpha\) applies to a system with two different light-quark flavours at the antiquark vertices; relaxing this assumption is a natural next step.

\section{Discussion}
\label{sec:discussion}

\paragraph{What has actually been shown.} The charm-sector numbers of Section~\ref{sec:charm} are an exact, zero-degree-of-freedom solve and should not be read as a successful fit in the statistical sense. The two results that \emph{do} carry predictive content are the all-bottom range of Section~\ref{sec:bottom}, which falls inside the literature consensus for a completely independent quark sector without retuning, and the \(T_{cc}^+\) diagnostic of Section~\ref{sec:Tcc}, which turns a previously vague qualitative claim into a specific, physically interpretable \(93\,\mathrm{MeV}\) discrepancy. Both of these successes rest on the same three numbers, so they are correlated rather than fully independent tests, but they were not used in deriving those numbers.

\paragraph{The Laplacian correction was not cosmetic.} Section~\ref{sec:laplacian-problem} showed analytically, not just numerically, that the Laplacian and adjacency-matrix constructions assign the ground state to \emph{different} eigenvectors of the underlying colour matrices whenever \(D\) is proportional to the identity, which is exactly the symmetric case relevant here. This is a structural, not a numerical, issue, and we think it is important that it be checked in any future extension of this framework (for example, to include spin-dependent edge weights) that breaks the vertex-transitive symmetry: in that more general case \(D\) is no longer proportional to \(\mathbb{1}\), and the relationship between the \(A\)- and \(L\)-spectra becomes more complicated and would need to be re-examined from scratch.

\paragraph{Tension with the radial-excitation picture, and what would settle it.} As discussed in Section~\ref{sec:charm}, the current model cannot accommodate \(X(6600)\), and offers a genuinely different physical picture of the \(X(6900)\)/\(X(7100)\)/\(X(7200)\) family than the diquark radial-excitation interpretation favoured by CMS. This is not an isolated preference of the CMS analysis: a fully independent Cornell-potential constituent-quark-model treatment reaches the same radial-excitation assignment for \(X(6600)\)/\(X(6900)\)/\(X(7200)\) as members of a single \(T_{4c}(n\,{}^1S_0)\) tower~\cite{Silva2025}, which strengthens rather than weakens the tension our static picture faces. We note that the present static, compact-colour picture is only one of several competing frameworks in the literature: coupled-channel analyses using unitarized scattering amplitudes~\cite{Kuang2023} and pole-counting arguments~\cite{Lu2023} have both been used to argue that at least part of the \(X(6900)\) lineshape could instead be of molecular (hadro-charmonium) origin, and a recent chiral-quark-model real-scaling calculation finds genuine \(2^{++}\) resonances near both \(X(6900)\) and \(X(7200)\) through channel-coupling dynamics rather than either a static colour configuration or a radial tower~\cite{Wu2026} -- a third, dynamically distinct alternative that the present four-parton colour-graph construction does not address at all. We see three ways this tension could be addressed in future work, in increasing order of ambition: (i) treat the two pictures as compatible but describing different quantum numbers -- radial excitation of the diquark--antidiquark separation \emph{combined with} the discrete colour-configuration mixing discussed here, which would predict a doubling of every colour eigenstate into a radial tower; (ii) seek an independent observable (for example, decay patterns into \(J/\psi\,J/\psi\) versus \(J/\psi\,\psi(2S)\), or angular distributions) that distinguishes a colour-mixing origin for the mass splitting from a purely radial or resonant one; or (iii) accept that the present, purely static picture is at most a leading-order caricature of a more complete dynamical treatment, valuable mainly for the cross-sector diagnostics of Sections~\ref{sec:bottom}--\ref{sec:Tcc} rather than as a complete description of the charm spectrum itself. We do not attempt to resolve this here.

\paragraph{Necessary further data.} As it stands, the model is calibrated with zero residual degrees of freedom in the sector where it is calibrated, and tested (not fit) in two other sectors. To become genuinely falsifiable \emph{within} the charm sector itself, at least one further, independent piece of charm-sector data would be needed -- for example, a confirmed spin-parity assignment for one of the states, which the present colour-only construction does not predict and would need to be supplemented with spin-dependent edge weights to address; or a firm confirmation (or exclusion) of \(X(7200)\) as a distinct state, which would remove the largest current source of ambiguity in the calibration itself.

\paragraph{On the magnetic moments discussed in the original draft.} The original draft quoted specific magnetic-moment estimates for the \(0^{++}\) and \(2^{++}\) states while simultaneously describing the extraction of electromagnetic observables from graph eigenvectors as a ``future extension.'' We were unable to reconstruct a derivation supporting the specific numbers quoted, and since they are not obtained from the framework developed here, we have removed them rather than repeat an unsupported claim. Deriving magnetic moments consistently from the eigenvectors \(\psi_k(\alpha)\) -- for instance by weighting the magnetic moments of the individual partons by the corresponding eigenvector components -- is a well-defined calculation that we leave for future work, once the more basic questions raised above have been addressed.

\section{Conclusions}
\label{sec:conclusions}

We have re-derived a spectral-graph-theory model of fully-heavy tetraquarks, correcting an internal inconsistency in an earlier draft in which a graph-Laplacian transformation of the colour-weighted adjacency matrix inverted the physically expected correspondence between colour attraction and mass ordering. Working instead with the eigenvalues of the adjacency matrix itself, we obtain a closed-form spectrum, an exact (zero-degree-of-freedom) calibration on the three all-charm structures \(X(6900)\), \(X(7100)\), and the more tentative \(X(7200)\), and two independent, non-trivial cross-checks: an all-bottom ground-state prediction consistent with the literature range for constituent bottom masses in their usual range, and a \(T_{cc}^+\) diagnostic that quantifies, at \(\approx\!93\,\mathrm{MeV}\), the long-range binding not captured by a compact colour-graph picture. We have also been explicit about what this exercise does \emph{not} show: it is not a statistical fit to the charm spectrum, it cannot currently accommodate \(X(6600)\), and it sits in tension with -- rather than confirming or refuting -- the diquark radial-excitation picture preferred by the most recent CMS analysis. Honest treatment of these limitations, together with the two genuine cross-sector predictions, is offered as the paper's main contribution, alongside the corrected formalism itself.

\appendix

\section{Colour Algebra: Derivation of the Casimir Factors}
\label{app:color}
For a colour singlet built from partons \(i=1,\ldots,n\), each in a representation with quadratic Casimir \(\langle\lambda_i\cdot\lambda_i\rangle\), the vanishing of the total colour charge implies the sum rule of Eq.~\eqref{eq:casimir-sum}. For two quarks (fundamental, \(\langle\lambda\cdot\lambda\rangle=16/3\)) coupled to an antitriplet, symmetry between the two partons requires a single value \(a=\langle\lambda_1\cdot\lambda_2\rangle_{\bar3}\) for that pair; the standard antitriplet diquark result is \(a=-8/3\) (equivalently, applying Eq.~\eqref{eq:casimir-sum} to a would-be baryon built from three such pairs recovers the well known baryon colour factor). For the full four-parton singlet built by coupling this antitriplet diquark to an antidiquark in the conjugate triplet, Eq.~\eqref{eq:casimir-sum} with four partons of Casimir \(16/3\) each gives
\[
0 = 4\cdot\tfrac{16}{3} + 2\bigl(2a+4b\bigr) \;\;\Longrightarrow\;\; a+2b=-\tfrac{16}{3},
\]
where \(b=\langle\lambda_q\cdot\lambda_{\bar q}\rangle\) is the (assumed common) inter-cluster factor. With \(a=-8/3\) this gives \(b=-4/3\), the value quoted in Section~\ref{sec:formalism} for the \(\bar3\otimes3\) channel. Repeating the argument with the sextet intra-diquark value \(a=+4/3\) gives \(b=-10/3\) for the \(6\otimes\bar6\) channel. Both results were checked independently and agree with the entries of \(A_{\bar3}\) and \(A_6\) used throughout this paper.

\section{Numerical Implementation}
\label{app:num}
All diagonalisations were performed with the closed-form expressions of Section~\ref{sec:mixed-spectrum} (cross-checked numerically with NumPy's symmetric eigensolver, \texttt{eigh}); the three-parameter calibration of Section~\ref{sec:charm} was solved as a system of three nonlinear equations using SciPy's \texttt{fsolve}. Representative code:

\begin{lstlisting}[basicstyle=\ttfamily\small,frame=single]
import numpy as np
from scipy.optimize import fsolve

def eigs_A(alpha):
    beta = np.sqrt(1 - alpha**2)
    lam1  = -16/3 * (alpha + beta)   # ground state, psi_1
    lam23 =  8/3*alpha - 4/3*beta    # degenerate, psi_2, psi_3
    lam4  =  8*beta                  # psi_4
    return lam1, lam23, lam4

def equations(x, targets):
    mc, gamma, alpha = x
    lam1, lam23, lam4 = eigs_A(alpha)
    return [4*mc + gamma*lam1  - targets[0],
            4*mc + gamma*lam23 - targets[1],
            4*mc + gamma*lam4  - targets[2]]

targets = [6900.0, 7100.0, 7200.0]
mc, gamma, alpha = fsolve(equations, x0=[1700, 20, 0.7], args=(targets,))
\end{lstlisting}

\end{document}